\documentclass[fleqn,usenatbib]{mnras}
\usepackage{newtxtext,newtxmath,anyfontsize}
\usepackage[T1]{fontenc}

\DeclareRobustCommand{\VAN}[3]{#2}
\let\VANthebibliography\thebibliography
\def\thebibliography{\DeclareRobustCommand{\VAN}[3]{##3}\VANthebibliography}

\usepackage{graphicx}	
\usepackage{amsmath}	

\newcommand{\rev}{}

\title[Surviving sublimation around hot WDs]{Where planetary solids survive sublimation around young and hot white dwarfs}

\author[]{Dimitri Veras$^{1,2,3}$\thanks{E-mail:dimitri.veras@aya.yale.edu},
Keiji Ohtsuki$^{4}$,
Rafael Martinez-Brunner$^{3}$,
Takato Nishio$^{4}$,
Ryo Tamon$^{4}$
\\
$^{1}$Centre for Exoplanets and Habitability, University of Warwick, Coventry CV4 7AL, UK
\\
$^{2}$Centre for Space Domain Awareness, University of Warwick, Coventry CV4 7AL, UK
\\
$^{3}$Department of Physics, University of Warwick, Coventry CV4 7AL, UK
\\
$^{4}$Department of Planetology, Kobe University, 1-1 Rokkodai, Nada, Kobe 657-8501, Japan
}

\pubyear{\the\year{}}

\begin{document}
\label{firstpage}
\pagerange{\pageref{firstpage}--\pageref{lastpage}}
\maketitle

\begin{abstract}
All observed planetary systems orbiting single white dwarfs have lived through the hot stellar transition from an asymptotic giant branch star. In this post-nebular transition period, the initial conditions for planetary system evolution throughout white dwarf cooling are established. The hottest ($\gtrsim 20$,$000$~K) and youngest ($\lesssim 20$~Myr-old) white dwarf planetary system host stars differ significantly from their canonical older and colder counterparts by failing to support solid body accumulation in the immediate vicinity (a few $R_{\odot}$) of the white dwarf. Here, we analyse the likely locations of both solid-body survival and the sublimated gaseous content during this pivotal epoch, and the consequences. We find that (i) reservoirs of iron-rich, rocky and water-rich asteroids of radius $R$ that later observably enrich, or pollute, the white dwarf need to remain parked for the first tens of Myr of white dwarf cooling beyond critical distances of (16~au)$\sqrt{1{\rm km}/R}$ (for iron), (30~au)$\sqrt{1{\rm km}/R}$ (for rock) and (130~au)$\sqrt{1{\rm km}/R}$ (for snow), (ii) sublimation acts much more quickly than radiatively-driven orbital drifts from Poynting-Robertson drag or the Yarkovsky effect, and (iii) although large asteroids ($R \approx 10-1$,$000$~km) that are kicked on highly eccentric orbits around newly born white dwarfs could survive sublimation, they may fragment into debris which will sublime before the white dwarf cools. These results support, but do not necessitate, dynamical origin scenarios of polluted white dwarfs that feature delayed gravitational instability subsequent to the host star's asymptotic giant branch phase at Kuiper Belt-like distances, and beyond.
\end{abstract}




\section{Introduction}

The course of evolution of a white dwarf planetary system is determined through the architecture established at the moment the white dwarf is born. It is born hot, with an effective temperature, $T_{\rm WD}$, as high as $10^5$~K. Then, within just 20~Myr, $T_{\rm WD}$ decreases by about 70-80~per\,cent. Within this timeframe, there have been discoveries of particularly unique systems such as the Helix nebula \citep{suetal2007}, the WD~J0914+1914 planetary system \citep{ganetal2019}, and the HS~0209+0832 planetary system (Williams, G\"{a}nsicke, et al., In Press 2026).

However, the dearth of discoveries during this hot post-nebular phase is perhaps even more notable. The vast majority of post-main-sequence planetary systems become observable later, after the white dwarfs have cooled down enough (below $\sim$ 20,000~K) to allow for dust and photospheric metal enrichment from rocky bodies to be seen \citep{couetal2019,manetal2020,wiletal2024,xuetal2024,nooetal2025,bonsor2026}. Evidence for gaseous and volatile-enhanced systems around these older white dwarfs do exist and are in the clear minority, but represent examples that have garnered a high level of interest \citep{faretal2013,radetal2015,xuetal2017,holetal2018,hosetal2020,putxu2021,johetal2022,sahetal2025,wiletal2025}.

Theoretical studies that are dedicated to  young ($\lesssim 20$~Myr-old) white dwarf planetary systems are also rare, and have focussed either on individual cases \citep{verful2020,veras2020,zotver2020,maretal2023}, natal kicks just after the white dwarf is born \citep{stoetal2015,ocoetal2023b,akietal2024,pharei2024}, or planetary evaporation \citep{villiv2007,schetal2019,galetal2024,becetal2025}. Many other studies consider the post-nebular, earliest white dwarf phase just in passing, often sampling it at a low time resolution in $N$-body simulations that incorporate stellar phase changes \citep[e.g.][]{debsig2002,vermusbonetal2013,musetal2014,maletal2020a,smaetal2021,baretal2022,vergeodob2023}.

\begin{figure*}
\includegraphics[width=15cm]{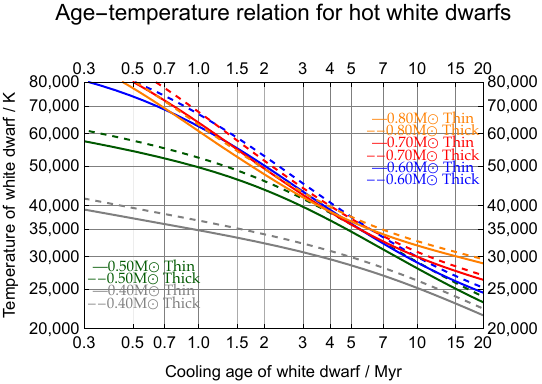}
\caption{
The relation between effective temperature, $T_{\rm WD}$, and cooling age, $t_{\rm cool}$ (the time since the star became a white dwarf), for hot white dwarfs of five different masses and two different envelope compositions \citep{bedetal2020}. The models decrease in accuracy for $T_{\rm WD}>80$,000K and $t_{\rm cool} < 0.3$~Myr, and hence these parameter space regions are not considered (also see https://warwick.ac.uk/fac/sci/physics/research/astro/people/antoinebedard/models/). The white dwarfs labelled ``thin'' correspond to an envelope with a hydrogen mass content equal to $10^{-10}M_{\rm WD}$ and a helium mass content equal to  $10^{-2}M_{\rm WD}$. The white dwarfs labelled ``thick'' correspond to an envelope with a hydrogen mass content equal to $10^{-4}M_{\rm WD}$ and a helium mass content equal to  $10^{-2}M_{\rm WD}$. This paper focusses on white dwarfs that are within the bounds of this plot.
}
\label{Fig:AgeTemp}
\end{figure*}

Here, we analyse the likely prevalence and distribution of solid-body survival and sublimated gas in young ($\lesssim$20~Myr) and hot ($T_{\rm WD}\gtrsim$~20,000~K) white dwarf planetary systems (see Fig. \ref{Fig:AgeTemp}). To do so, we justify decoupling treatments of sublimative forces and radiative drag, and focus on the region exterior to the Roche radius of the white dwarf\footnote{A white dwarf's Roche radius approximately equals $1R_{\odot}$ for a variety of assumptions; for gravity-dominated objects, some assumptions are outlined in Table 1 of \cite{vercarleietal2017}. For strength-dominated objects, see Eq. (A.2) of \cite{zhoetal2024}. A comparison of different prescriptions can be found in the Appendix of \cite{steetal2026}. In our order-of-magnitude study, we simply assume that the Roche radius always equals $1R_{\odot}$.}, within which debris discs are commonly modelled. In fact, around these hot white dwarfs, the characteristic sublimation radius exceeds the Roche limit \citep{steckloffetal2021a}, suggesting that the Roche sphere does not play a significant role here (unlike for most of the compact white dwarf debris discs which are common around older white dwarfs; \citealt{nooetal2025,malamud2026}).

To perform our study, we employ a combination of basic physical prescriptions from \cite{vereggganGB2015}, \cite{mcdver2021} and \cite{verbirzam2022}. 
\cite{vereggganGB2015} presented unaveraged and averaged equations for the effect of arbitrary giant branch radiative forces on particles, including those from Poynting-Robertson drag, radiation pressure, and the Yarkovsky effect. These relations also apply around white dwarfs. \cite{mcdver2021} extended the work of \cite{broetal2017} and modelled the fragmentation and sublimation of triaxial asteroids, boulders and smaller objects on direct collision courses with white dwarfs of different temperatures. \cite{verbirzam2022} modelled the effect of Poynting-Robertson drag, the Yarkovsky effect, and the YORP effect on dust, pebbles, boulders and asteroids around white dwarfs of all ages, but did not consider sublimation.

Because we do not require a complex sublimation prescription, we adopt the rough estimate used in \cite{mcdver2021}. 
We acknowledge that in white dwarf planetary systems, the process of sublimation has been widely studied in a variety of other contexts. These include within white dwarf debris discs \citep{rafgar2012,metetal2012,sheetal2019,treetal2021,steckloffetal2021a,okuetal2023,froetal2024}, just outside of these discs \citep{vereggganSUB2015,sheser2022,verassteckloff2026}, and in the wider vicinity \citep{stoetal2015,zhoetal2024}.

In Section 2, we lay out analytical prescriptions for the aforementioned forces, and derive relations that are useful for this study. We then compute the results in Section 3, discuss them in Section 4 and conclude in Section 5.

\section{Force prescriptions}

In order to compare sublimative and radiative forces exterior to the white dwarf Roche radius, we provide analytical prescriptions for each\footnote{The effects of ablation and space weathering should be negligible. White dwarfs have no winds, and so, unlike for giant branch stars, the interplanetary medium is relatively barren.}. The most relevant radiative forces are Poynting-Robertson drag, radiation pressure, the Yarkovsky effect and the YORP effect.

Each effect is most relevant in different size regimes. \cite{verbirzam2022} suggest that Poynting-Robertson drag primarily acts on objects with radii $R=10^{-5}-10^0$~m, the Yarkovsky effect primarily acts on objects with radii $R=10^{-1}-10^6$~m and the YORP effect primarily acts on objects with radii $R=10^0-10^6$~m. 

Another potentially important forcing is that due to SYORP, or the sublimative-YORP effect \citep{stejac2016,safetal2021}. Rotational fission due to SYORP has been shown to occur more quickly than from YORP \citep{verassteckloff2026}. However, both the YORP and SYORP effects only apply for aspherical objects, and is highly shape-dependent. These effects represent additional methods to break apart asteroids in addition to tidal disruption or collisional disruption, but do not alter orbits. {\rev We compare SYORP and YORP timescales to the breakup timescales we compute here in Section 4.}

For the subsequent expressions, rather than using stellar luminosity ($L_{\rm WD}$), we instead use the more direct observable temperature ($T_{\rm WD}$). These quantities are related through
~
\begin{equation}
L_{\rm WD} = 4 \pi R_{\rm WD}^2 \sigma T_{\rm WD}^4,
\end{equation}

\noindent{}where $R_{\rm WD}$ is the radius of the white dwarf and $\sigma$ is the Stefan-Boltzmann constant. 

\subsection{Assumptions}

{\rev In both Sections 2.2 and 2.3, we adopt a common set of assumptions. The first is -- because we are neglecting the YORP and SYORP effects -- that the sublimating body is spherical with a radius $R$. The next assumptions are that the object is composed of the same material throughout its radius, the object is efficient at converting all of the incident radiation into heat that can generate submlimation, and the object is immune to more complex processes such as cooling or the effect of intrinsic vapour pressure; here we seek just order-of-magnitude results, as is appropriate given the available data.} 

{\rev Additionally, because we are considering a coupled treatment of sublimation and orbital shifts, the variables $a$, $e$, $\omega$ and $f$ -- which represent the object's semi-major axis, eccentricity, true anomaly and argument of pericentre -- as well as $R$, are always a function of time in our equations. We explicitly illustrate that by including a time dependence on those variables. However, in the {\it derivation} of most of the averaged equations (which has been done elsewhere and is not the focus of this paper), $a$ and $e$ are often treated as constant, an assumption that is justified in a previous coupled treatment \citep{vereggganSUB2015}. This assumption implies that these variables do not change significantly over the course of a single orbit.}

{\rev Further, the magnitude of the Yarkovsky effect is in particular sensitive to one's assumptions about the components of the object's spin angular momentum vector and orbital angular momentum vector, and how both change with time \citep{vereggganGB2015}. For our purposes here, we need only consider the maximum possible effect, as we show that the Yarkovsky effect is negligible compared to the speed of sublimation.}

\subsection{Unaveraged equations}

The equations in this subsection can be used to trace orbital and physical changes in an object due to radiation and sublimation along a single orbit.

\subsubsection{Poynting-Robertson drag and radiation pressure (for $R=10^{-5}-10^0$~m)}

In a {\rev single-star} system, the extra acceleration on orbiting dust or pebbles due to impinging radiation gives rise to two effects: Poynting-Robertson drag and radiation pressure. While the definitions of each term have been historically ambiguous, as discussed in \cite{buretal1979}, currently radiation pressure is assumed to be an outward force and Poynting-Robertson drag is assumed to be an inward force.

Both forces can be expressed together as a single acceleration term in the equations of motion. Based on \cite{buretal2014}, Eqs. (5-7) of \cite{vereggganGB2015} read:
~
\begin{equation}
\left(\frac{d^2r}{dt^2}\right)_{\rm PR+RP}
=
\frac
{3 \sigma R_{\rm WD}^2 T_{\rm WD}^4 \left(Q_{\rm abs} + Q_{\rm ref} \right)}
{4 c R(t) \rho r^2}
\left[
\left(1 - \frac{\vec{v}\cdot\vec{r}}{cr}  \right)\frac{\vec{r}}{r}
-\frac{\vec{v}}{c}
\right],
\label{UnavPRRP}
\end{equation}

\noindent{}where $Q_{\rm abs}$ and $Q_{\rm ref}$ are the object's absorption efficiency and reflecting efficiency, respectively. Both of these parameters are dimensionless constants between 0-1. The object's density is $\rho$, the speed of light is $c$, and the object's distance and velocity with respect to the centre of the white dwarf are given by $ \vec{r}$ and $\vec{v}$.

\subsubsection{The Yarkovsky effect (for $R=10^{-1}-10^6$~m)}

The complete expressions for the orbital element evolution due to the Yarkovsky effect \citep{vereggganGB2015} are unnecessarily complex for our applications here, particularly because we will demonstrate that the maximum possible averaged expression is negligible compared to the effects of sublimation.

\subsubsection{Sublimation}

In this case, we assume that sublimation gradually reduces the object's mass $M$ until nothing is left. 
Consequently, Eq. (A2) of \cite{mcdver2021} gives
~
\begin{equation}
\frac{dM(t)}{dr}
=
-
\frac
{\pi \sigma R(t)^2 T_{\rm WD}^4}
{\mathcal{L} v(t) \left( \frac{r(t)}{R_{\rm WD}} \right)^2},
\label{SubMR}
\end{equation}

\noindent{}where $\mathcal{L}$ is the latent heat of the object. \cite{mcdver2021} give values of $\mathcal{L}=2.6\times 10^6$~J/kg and $\rho=0.5$~g/cm$^3$ for snow, $\mathcal{L}=8\times 10^6$~J/kg and $\rho=3$~g/cm$^3$ for rock, and $\mathcal{L}=1\times 10^7$~J/kg and $\rho=8$~g/cm$^3$  for iron. By incorporating this time dependence in equation (\ref{SubMR}), we avoid the commonly-adopted notion of a single sublimation radius for a given material.

It follows that
~
\begin{equation}
\frac{dR(t)}{dr}
=
-
\frac
{\sigma R_{\rm WD}^2 T_{\rm WD}^4}
{4 \rho \mathcal{L} v(t) r(t)^2}
\end{equation}

\noindent{}and so
~
\begin{equation}
\frac{dR(t)}{dt}
=
-
\frac
{\sigma R_{\rm WD}^2 T_{\rm WD}^4}
{4 \rho \mathcal{L} r(t)^2}
=
-
\frac
{\sigma R_{\rm WD}^2 T_{\rm WD}^4 \left(1 + e(t) \cos{f(t)} \right)^2}
{4 \rho \mathcal{L} a(t)^2 \left( 1 - e(t)^2 \right)^2}
.
\label{drdt}
\end{equation}

\begin{figure*}
\includegraphics[width=15cm]{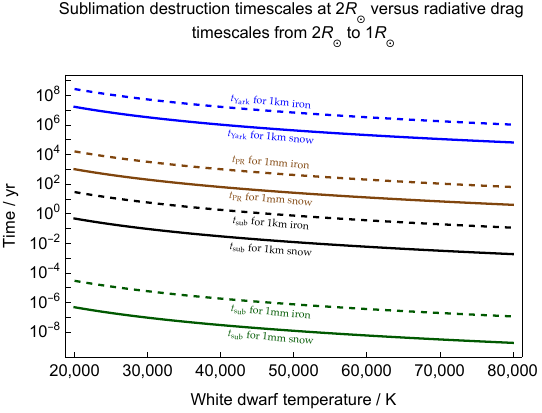}
\caption{
Comparison of the orbit-averaged sublimiative destruction timescales ($t_{\rm sub}$) and orbit-averaged radiative drag timescales ($t_{\rm PR}$ and $t_{\rm Yark}$) for a 1~km asteroid (blue and black curves) and a 1~mm pebble (brown and green curves) each of uniform composition (equations \ref{tPRmax}-\ref{tdest}). Compositions of iron are indicated by dashed curves and of snow are indicted by solid curves. The sublimation all takes place at a separation of $2R_{\odot}$ from a $0.6M_{\rm WD}$ white dwarf, and the radiative migration occurs from a semi-major axis of $2R_{\odot}$ to $1R_{\odot}$. The plot demonstrates that sublimation occurs quickly enough that the process is effectively decoupled from radiative drag, which is rendered to be largely ineffectual beyond the white dwarf Roche radius unless the sublimation is suppressed by effects that are not modelled here.
}
\label{Fig:tComp}
\end{figure*}

\subsection{Averaged equations}

The equations in this subsection can be used to trace orbital and physical changes in an object over many orbits.

\subsubsection{Poynting-Robertson drag and radiation pressure (for $R=10^{-5}-10^0$~m)}

{\rev The acceleration expressed in Eq. (\ref{UnavPRRP}) may be used to obtain averaged equations of motion for orbital elements only if the effects of radiation pressure and Poynting-Robertson drag are negligible over the course of a single orbit. Because radiation pressure is stronger than Poynting-Robertson drag, we need only establish the strength of radiation pressure.}

{\rev To do so, we} consider the {\rev widely-used $\beta$ formulation \citep{buretal1979} for radiation pressure}: e.g., Eq. (5) and Fig. 10 of \cite{rigwya2022} show that any grains with $\beta>1$ will be radiatively blown out of the system, where:
\[
\beta = \frac
{3 \left(Q_{\rm abs} + Q_{\rm ref} \right) \sigma R_{\rm WD}^2 T_{\rm WD}^4}
{4 G c M_{\rm WD} R(t) \rho}
\]
\vspace{-4mm}
\[
\ \ \ = 0.98 
\left( \frac{T_{\rm WD}}{50{\rm ,}000\,{\rm K}} \right)^{4}
\left( \frac{R(t)}{1\,\mu{\rm m}} \right)^{-1}
\left( \frac{M_{\rm WD}}{0.4M_{\odot}} \right)^{-1}
\]
\vspace{-4mm}
\begin{equation}
\ \ \ \ \ \ \ \ \ \ \ \ 
\times 
\left( \frac{R_{\rm WD}}{0.016R_{\odot}} \right)^{2}
\left( \frac{\rho}{2 {\rm g/cm}^3} 
\right)^{-1},
\label{blowout}
\end{equation}
\noindent{}where we have deliberately chosen extreme values for $T_{\rm WD}$, $R$, $M_{\rm WD}$ and $R_{\rm WD}$. Eq. (\ref{blowout}) shows that even for relatively light white dwarfs\footnote{The vast majority of single white dwarfs have masses in the range $M_{\rm WD}=0.4-0.8M_{\odot}$  \citep[e.g.][]{cumetal2018,elbetal2018,obretal2024}.}, micron-sized dust can only be pushed out when $T_{\rm WD}$ exceeds 50,000~K. For all applications here, micron-sized dust is too small to withstand sublimation before being blown out of the system.  
Hence, effectively, once an object becomes trapped in the gravitational well of a white dwarf, we can assume {\rev that radiation pressure produces a negligible change to orbital parameters over the course of one orbit. Further, because white dwarfs are too dense to allow for stellar winds (an important consideration for exozodiacal dust and debris discs around main-sequence stars), we do not need to consider the possibility of winds pushing grains out of the system.}

{\rev Consequently, we can utilise the averaged equations of motion based on Eq. (\ref{UnavPRRP}).} These equations have been widely used for decades in investigations of dust dynamics. They are based on \cite{wyawhi1950}; from Eqs. (1-2) of \cite{verbirzam2022}:
~
\begin{equation}
\left\langle \frac{da(t)}{dt} \right\rangle_{\rm PR}
=
-\frac
{3 \left(Q_{\rm abs} + Q_{\rm ref} \right)\left(2+3e(t)^2\right)\sigma R_{\rm WD}^2 T_{\rm WD}^4}
{4c^2R(t)\rho a(t) \left(1-e(t)^2\right)^{3/2}},
\label{PRaAvg}
\end{equation}
~
\begin{equation}
\left\langle \frac{de(t)}{dt} \right\rangle_{\rm PR}
=
-\frac
{15 \left(Q_{\rm abs} + Q_{\rm ref} \right)e(t)\sigma R_{\rm WD}^2 T_{\rm WD}^4}
{8c^2R(t)\rho a(t)^2 \sqrt{1-e(t)^2}},
\end{equation}
~
\begin{equation}
\left\langle \frac{d\omega(t)}{dt} \right\rangle_{\rm PR}
=
0,
\end{equation}

\noindent{}where $\omega$ represents the osculating  argument of pericentre of the object's orbit. Hence, the evolution of the orbital pericentre, $q(t)=a(t)[1-e(t)]$, is
~
\begin{equation}
\left\langle \frac{dq(t)}{dt} \right\rangle_{\rm PR}
=
-\frac
{3 \left(Q_{\rm abs} + Q_{\rm ref} \right)\sigma R_{\rm WD}^2 T_{\rm WD}^4 \sqrt{1-e(t)} \left(4-e(t) \right)}
{8c^2R(t)\rho a(t) \left(1+e(t)\right)^{3/2}}.
\label{PRqAvg}
\end{equation}

\subsubsection{The Yarkovsky effect (for $R=10^{-1}-10^6$~m)}

Unlike for Poynting-Robertson drag, the averaged equations of motion for the Yarkovsky effect are usually restricted to $\langle da(t)/dt \rangle$ and have been applied almost exclusively within the solar system \citep[e.g.][]{voketal2015}. The Sun's current luminosity is too small to typically justify modelling changes in other orbital elements. 

However, in systems where the star is much more luminous than the sun, at least $\langle de(t)/dt \rangle$ may be non-negligible. \cite{vereggganGB2015}, \cite{verhigida2019} and \cite{feretal2022} computed the effect for giant branch star planetary systems, whose stars can be 3-4 orders of magnitude more luminous than the sun. Because white dwarfs which are younger than a few Myr have luminosities which exceed that of the Sun, here we also consider $\langle de(t)/dt \rangle$, and write $\langle da(t)/dt \rangle$ as a function of $e(t)$. Further, we consider only the maximum possible value of these rates, in order to allow the averaging to be analytically tractable and to be conservative, as we will show that the effects are relatively negligible.

These maximum rates are given by Eqs. (3-4) of \cite{verbirzam2022} as:
~
\begin{equation}
\left\langle \frac{da(t)}{dt} \right\rangle_{\rm Yark,\ Max}
\approx
-\frac
{3 \sigma R_{\rm WD}^2 T_{\rm WD}^4}
{16 c R(t)\rho \sqrt{G M_{\rm WD} a(t)}\left(1-e(t)^2  \right)},
\label{YarkaAvg}
\end{equation}
~
\[
\left\langle \frac{de(t)}{dt} \right\rangle_{\rm Yark,\ Max}
\approx
-\frac
{3 \sigma R_{\rm WD}^2 T_{\rm WD}^4}
{32 c R(t)\rho \sqrt{G M_{\rm WD} a(t)^3}} 
\]
~
\begin{equation}
\ \ \ \ \ \ \ \ \ 
\ \ \ \ \ \ \ \ \ 
\ \ \ \ \ \, 
\times \left[ \frac{e(t)^2 - 2\left(1-e(t)^2\right)\left(1 - \sqrt{1-e(t)^2} \right)}{e(t)^3} \right],
\end{equation}
\noindent{}and so
~
\[
\left\langle \frac{dq(t)}{dt} \right\rangle_{\rm Yark,\ Max}
\approx
\frac
{3 \sigma R_{\rm WD}^2 T_{\rm WD}^4}
{32 c R(t)\rho \sqrt{G M_{\rm WD} a(t)}} 
\]
~
\begin{equation}
\ \ \ \ \ \ \ \ \ 
\ \ \ \ \ \ \ \ \ 
\ \ \ \ \ \, 
\times \left[ \frac{e(t)^2 - 2\left(1-e(t)^2\right)\left(1 - \sqrt{1-e(t)^2} \right)}{e(t)^3} - \frac{2}{1+e(t)}\right].
\label{YarkqAvg}
\end{equation}
\noindent{}The expression for $\langle d\omega(t)/dt \rangle$ is nonzero, but is long and difficult to compute (see Eq. A6 of \citealt{vereggganGB2015}), and is anyway unnecessary as we will show that the Yarkovsky effect is negligible compared to sublimation in hot white dwarf systems. 

There is a small potential radius range of overlap ($R=10^{-1}-10^0$~m) in which both the Yarkovsky effect and Poynting-Robertson drag may be important \citep{veras2020}. The Yarkovsky effect is a factor of $\sim(c/v_{\rm K})$ stronger (where $v_{\rm K} = \sqrt{GM_{\rm WD}/a(t)}$), but whether the Yarkovsky effect activates depends on the material properties and geometry of the object.

\subsubsection{Sublimation}

In order to make a direct comparison with radiative drag, we need to obtain an expression for the orbit-averaged sublimated material. We average Eq. (\ref{drdt}) with
~
\[
\left\langle 
\frac{dR(t)}{dt}
\right\rangle_{\rm sub}
=
\frac{n}{2\pi}
\int_{0}^{2\pi}
\frac{dR(t)}{dt}
\frac{dt}{df}
df
\]
~
\[
\ \ \ \ \ \ \
\ \ \ \ 
=
\frac{n}{2\pi}
\int_{0}^{2\pi}
\frac{dR(t)}{dt}
\frac{\left(1 - e^2\right)^{3/2}}{n\left(1 + e \cos{f} \right)^2}
df
\]
~
\begin{equation}
\ \ \ \ \ \ \
\ \ \ \ 
=
-\frac
{\sigma R_{\rm WD}^2 T_{\rm WD}^4}
{4 \rho \mathcal{L} a(t)^2 \sqrt{1-e(t)^2}},
\label{drdtAvg}
\end{equation}

\noindent{}where $n$ represents the object's mean motion. This computation assumes that $a$ and $e$ remain constant over a single orbital period.

\section{Results}

Now we consolidate these relations in order to obtain our results. We report these results first with a comparison between radiative and sublimative effects (Section 3.1) before exploring the parameter space of sublimative effects (Section 3.2), identifying the critical distance beyond which objects need to remain parked at the start of the white dwarf phase to be safe from sublimation (Section 3.3), and linking sublimative effects to fragmentation (Section 3.4).

\begin{figure}
\includegraphics[width=8.5cm]{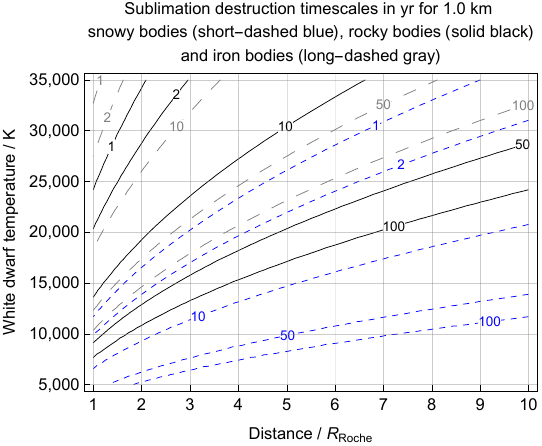}
\caption{
Approximate time for objects to sublimate completely, given as contour labels in {\rev 1, 2, 10, 50, 100 yrs}, computed from Eq. (\ref{drdtAvg}). {\rev The objects are 1~km homogeneous asteroids composed of snow (short-dashed blue), rock (solid black) and iron (long-dashed gray)}. In the computations, we assume circular orbits and  $R_{\rm WD} = 0.013R_{\odot}$, which corresponds to $M_{\rm WD}=0.6M_{\odot}$. The plots demonstrate that most objects under 1~km in size around hot white dwarfs completely sublimate within a few years within a few white dwarf Roche radii.
}
\label{Fig:Cont1}
\end{figure}

\begin{figure}
\includegraphics[width=8.5cm]{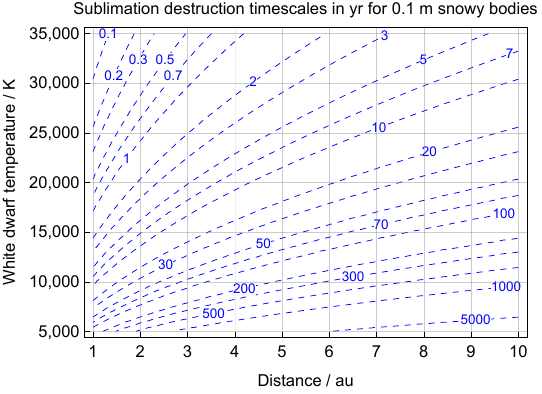}
\caption{
Like Fig. \ref{Fig:Cont1}, except for {\rev 0.1~m} objects and {\rev a spatial scale extending to 10~au}. Snowy boulders and pebbles sublimate into gas within a decade out to 10~au around hot white dwarfs.
}
\label{Fig:Cont2}
\end{figure}

\begin{figure*}
\includegraphics[width=15cm]{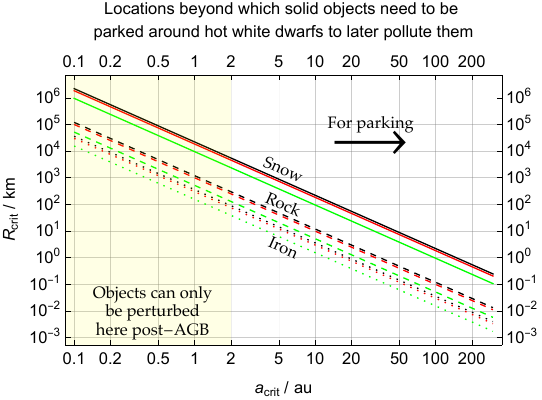}
\caption{
Value pairs of object radius and semi-major axis for which $t_{\rm cool} = t_{\rm sub}(T_{\rm WD}(t_{\rm cool}))$, for $(t_{\rm cool}, T_{\rm WD}) =$ $(0.3~{\rm Myr}, \ 88$,$812~{\rm K})$ (black curves), $(2~{\rm Myr}, \ 52$,$795~{\rm K})$ (red curves) and $(20~{\rm Myr}, \ 25$,$196~{\rm K})$ (green curves). Objects to the left of the curves are unlikely to survive until the white dwarf cools, meaning that such objects need to be parked sufficiently far away in order to later observably enrich, or pollute, the white dwarf. The coincidental overlap of the curves of similar compositions leads to the empirical estimates in Eqs. (\ref{SnowCrit})-(\ref{IronCrit}). The highlighted yellow region indicates a typical engulfment radius for a $M_{\rm WD}=0.6 M_{\odot}$ star, a value which is assumed in the computation of these curves for the thin model of \citet{bedetal2020}.
}
\label{Fig:Punch}
\end{figure*}

\subsection{Sublimation dominates radiative drifts}

Does radiation drag an object close to a hot white dwarf before the star sublimates the object? In this subsection, we show that the answer to this question is effectively ``no, not even close''.

To make this comparison, we first analyse how the object's changing radius affects the drift. Consider the averaged pericentre drift due to Poynting-Robertson drag and the Yarkovsky effect. In both Eqs. (\ref{PRqAvg}) and (\ref{YarkqAvg}), the drift rates are zeroed out in the limit $e(t)\rightarrow 1$. In the opposite extreme, with $e(t)\rightarrow 0$, the drift rates are maximised. 

{\rev In this limit, comparison to the averaged sublimation rate (Eq. \ref{drdtAvg}) yields, for Poynting-Robertson drag,}
~
\begin{equation}
\left\langle 
\frac{da(t)}{dt}
\right\rangle_{\rm PR}\Bigg|_{e(t)=0}
\ \Bigg/ \
\left\langle 
\frac{dR(t)}{dt}
\right\rangle_{\rm sub}
=
\frac{6\mathcal{L} \left(Q_{\rm abs} + Q_{\rm ref} \right)a(t)}{c^2 R(t)}
\end{equation}
{\rev which, upon integrating, gives}
\begin{equation}
\left( \frac{a_{\rm f}}{a_{\rm i}} \right)_{\rm PR}
=
\left( \frac{R_{\rm f}}{R_{\rm i}} \right)^
{\frac{6\mathcal{L} \left(Q_{\rm abs} + Q_{\rm ref} \right)}{c^2}} 
\sim 
\left( \frac{R_{\rm f}}{R_{\rm i}} \right)^
{2-7\times10^{-10}} \sim 1
\label{PRdep}
\end{equation}

\noindent{}{\rev for $\mathcal{L} = 2.6-10.0 \times 10^6$J/kg and $\left(Q_{\rm abs} + Q_{\rm ref} \right) = 1$.}

{\rev For the Yarkovsky effect,}
\begin{equation}
\left\langle 
\frac{da(t)}{dt}
\right\rangle_{\rm Yark}\Bigg|_{e(t)=0}
\ \Bigg/ \
\left\langle 
\frac{dR(t)}{dt}
\right\rangle_{\rm sub}
=
\frac{3\mathcal{L} a(t)^2}{4c R(t)\sqrt{G M_{\rm WD} a(t)}}
\end{equation}
{\rev which, upon integrating, gives}
~
\[
\left( \frac{a_{\rm f}}{a_{\rm i}} \right)_{\rm Yark}
=
\left[
1 - \frac{3\mathcal{L}}{8c}
\left(\frac{GM_{\rm WD}}{a_{\rm i}} \right)^{-1/2} \ln{\left(\frac{R_{\rm f}}{R_{\rm i}}\right)} 
\right]^{-2}
\]
~
\begin{equation}
\ \ \ \ \ \ \ \  
\ \ \ \ \ \ \ \
\sim \left[
1 - \left(7\times 10^{-9}-5\times10^{-6}\right) \ln{\left(\frac{R_{\rm f}}{R_{\rm i}}\right)} 
\right]^{-2}
\sim 1
\label{Yarkdep}
\end{equation}
\noindent{}{\rev for $\mathcal{L} = 2.6-10.0 \times 10^6$J/kg, $M_{\rm WD}=0.6M_{\odot}$ and $a_{\rm i} = 1R_{\odot}-100 {\rm au}$.}
The subscripts ``i'' and ``f'' indicate initial and final.

Eqs. (\ref{PRdep}) and (\ref{Yarkdep}) demonstrate that the radiative drift rate is effectively independent of the sublimative radius changes. Now, we consider the timescales for each process to act in isolation. In the limiting case of circular orbits, from Eq. (\ref{PRaAvg}) the Poynting-Robertson drag timescale is approximately
~
\begin{equation}
t_{\rm PR} \approx 
\left(a_{\rm i}^2 - a_{\rm f}^2 \right)
\left[
\frac{c^2 R \rho}
{3 \sigma R_{\rm WD}^2 T_{\rm WD}^4}
\right]
\label{tPRmax}
\end{equation}
\noindent{}and the (minimum) Yarkovsky drift timescale from Eq. (\ref{YarkaAvg}) is approximately 
~
\begin{equation}
t_{\rm Yark} \approx 
\left(a_{\rm i}^{3/2} - a_{\rm f}^{3/2} \right)
\left[
\frac{32 c R \rho \sqrt{G M_{\rm WD}}}
{9\sigma R_{\rm WD}^2 T_{\rm WD}^4}
\right].
\end{equation}
\noindent{}To obtain the timescale for an object to be sublimated completely, we integrate Eq. (\ref{drdtAvg}) and obtain
\begin{equation}
t_{\rm sub}
=
\frac
{4 R_{\rm i} \rho \mathcal{L} a^2 \sqrt{1-e^2}}
{\sigma R_{\rm WD}^2 T_{\rm WD}^4}.
\label{tdest}
\end{equation}

Equations (\ref{tPRmax})-(\ref{tdest}) reveal an equivalent but sensitive dependence of these timescales on the value of $T_{\rm WD}$; the timescales will decrease by a factor of $\approx$250 if $T_{\rm WD}$ is 80,000~K instead of 20,000~K. The timescales are also sensitive to the value of $R_{\rm WD}$. This value can be linked to $M_{\rm WD}$ through \citep{nauenberg1972}
~
\begin{equation}
\frac{R_{\rm WD}}{R_{\odot}}
\approx
0.0127 \left( \frac{M_{\rm WD}}{M_{\odot}} \right)^{-1/3}
\sqrt{1-0.607\left( \frac{M_{\rm WD}}{M_{\odot}} \right)^{4/3}}.
\label{Eq:MR}
\end{equation}
\noindent{}Hence, increasing $M_{\rm WD}$ from $0.4M_{\odot}$ to $0.8M_{\odot}$ will decrease the white dwarf's radius by about 35~per\,cent, and increase the timescales in equations (\ref{tPRmax})-(\ref{tdest}) by a factor of about 2.4.

In order to compare these three timescales, we now plot $t_{\rm PR}$, $t_{\rm Yark}$ and $t_{\rm sub}$ together in Fig. \ref{Fig:tComp}. We consider two object sizes -- an asteroid with radius 1~km and a pebble with radius 1~mm -- and two compositions, that of snow and iron. We also adopt $M_{\rm WD} = 0.6M_{\odot}$, which gives, through Eq. (\ref{Eq:MR}), $R_{\rm WD} = 0.013R_{\odot}$. In order to consider an extreme case, we compute the timescale for radiation to drag the object from a semi-major axis of $2R_{\odot}$ to $1R_{\odot}$, and assume that the sublimation {\rev takes place at a separation of $2R_{\odot}$}. 

The figure demonstrates that $t_{\rm sub} \ll t_{\rm PR}$ and $t_{\rm sub} \ll t_{\rm Yark}$. Further, the timescale differences are severe enough that altering the variable values in Eqs. (\ref{tPRmax})-(\ref{tdest}) will not change the final result. We henceforth decouple the processes of sublimation and radiative drag, and focus on the former. 

\begin{figure*}
\includegraphics[width=14cm]{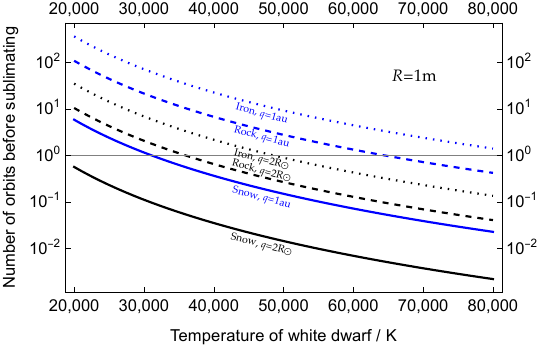}
\centerline{}
\centerline{}
\centerline{}
\centerline{}
\includegraphics[width=14cm]{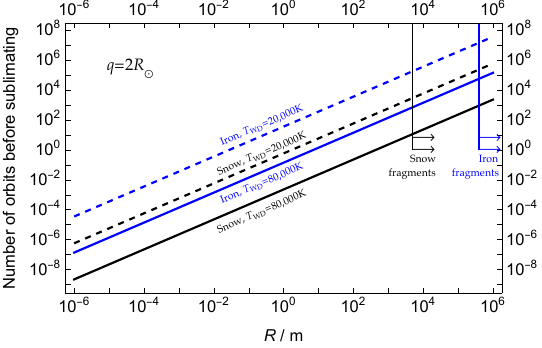}
\caption{
Number of orbits to sublime. This value, for highly eccentric orbits, is very weakly dependent on $a$ (Eq. \ref{PerEq}), which is taken here to be 100~au. Also, we assume $M_{\rm WD}=0.6M_{\odot}$, which sets a single orbital period to last 1,290~yr. The top panel curves all assume $R=1$~m, and the plot highlights the dependence on pericentre $q$ and white dwarf temperature. The bottom panel curves instead all assume $q=2R_{\odot}$, and the plot highlights the dependence on radius and white dwarf temperature. In this panel, the vertical lines indicate the maximum object radius that could withstand fragmentation at the orbital pericentre (Eq. \ref{FragLim}). The plot demonstrates that there does exist a range of parameter space that can support the survival of an asteroid on a highly eccentric orbit with a pericentre very close to the white dwarf's Roche radius throughout the hottest white dwarf epochs.
}
\label{Fig:NumOrbits}
\end{figure*}

\subsection{Parameter space exploration of sublimation}

The sublimation timescales plotted in Fig. \ref{Fig:tComp} suggest that gaseous remnants of planetary debris are widespread around hot white dwarfs. In this subsection, we quantify this notion through a parameter exploration study.

In Figs. \ref{Fig:Cont1}-\ref{Fig:Cont2}, we plot Eq. (\ref{drdtAvg}) to demonstrate how quickly objects sublime throughout the parameter space. Fig. \ref{Fig:Cont1} illustrates the result for 1~km asteroids composed purely of snow, rock or iron, all with the same radial extent (out to $10R_{\rm Roche}$). Fig. \ref{Fig:Cont2} {\rev focusses on a much further out region (out to 10 au) for smaller snowy objects.}

Overall, the destruction timescales shown on the plots demonstrate that hot white dwarfs cannot sustain surrounding rocky material for long. Within a few white dwarf Roche radii, most asteroids -- even composed entirely of iron -- that are smaller than 1~km will be sublimed within a few years. Snowy objects quickly become gaseous on au-scales.

\subsection{The critical parking distance}

The environment surrounding hot white dwarfs is clearly inhospitable for objects ranging from dust to pebbles to asteroids existing in solid form. This subsection addresses the related important question of just where does all of the solid material that we eventually observe around older white dwarfs reside? This material needs to be sufficiently distant from newly formed white dwarfs -- and stay there, parked -- for a sufficiently long time.

In order to estimate this critical parking distance, we first consider that a white dwarf's effective temperature decreases monotonically with cooling age, $t_{\rm cool}$, and hence $t_{\rm sub}$ increases with $t_{\rm cool}$. Then, by setting $t_{\rm cool} = t_{\rm sub}(T_{\rm WD}(t_{\rm cool}))$ and solving for $a$ and $R$, we obtain critical values denoted by $a_{\rm crit}(t_{\rm cool})$ and $R_{\rm crit}(t_{\rm cool})$. Coincidentally\footnote{The coincidence arises because for the $0.6M_{\odot}$ ``thin'' model in Fig.~\ref{Fig:AgeTemp}, $T_{\rm WD} \sim$ 63,000K $(t_{\rm cool}/{\rm Myr})^{-0.3}$. Hence, from equation (\ref{tdest}), $t_{\rm sub} \propto R_{\rm i}a^2 t_{\rm cool}^{1.2}$. As a result, solving $t_{\rm cool} = t_{\rm sub}(T_{\rm WD}(t_{\rm cool}))$ gives $R_{\rm i}a^2 \propto t_{\rm cool}^{-0.2}$, which demonstrates the weak dependence ($R_{\rm i}a^2$ varies by just a factor of about $2$ when $t_{\rm cool}$ varies between 0.3~Myr and 20~Myr).}, the functional form of Eq. (\ref{drdtAvg}) and the dependence of $t_{\rm cool}$ on $T_{\rm WD}$ helpfully combine in a way such that these critical values become largely independent of $t_{\rm cool}$, such that $a_{\rm crit} \approx a_{\rm crit}(t_{\rm cool})$ and $R_{\rm crit} \approx R_{\rm crit}(t_{\rm cool})$.

We plot $a_{\rm crit}$ and $R_{\rm crit}$ in Fig. \ref{Fig:Punch} for a $M_{\rm WD}=0.6M_{\odot}$ ``thin'' white dwarf. We plot curves for three different pairs of $(t_{\rm cool}, T_{\rm WD})$ values -- extrapolated from the grids in \cite{bedetal2020} -- as black $(0.3~{\rm Myr}, \ 88$,$812~{\rm K})$, blue $(2~{\rm Myr}, \ 52$,$795~{\rm K})$ and green $(20~{\rm Myr}, \ 25$,$196~{\rm K})$ lines. 

On the scale of this plot, within half of an order of magnitude of $R_{\rm crit}$, the coloured lines conveniently overlap with one another. This correspondence allows us to estimate a single relation between $R_{\rm crit}$ and $a_{\rm crit}$ for each composition studied here. These are
~
\begin{equation}
{\rm (SNOW)} \   R_{\rm crit} \approx 16,500~{\rm km} 
\left(\frac{a_{\rm crit}}{{\rm au}}\right)^{-2}
\ \ {\rm or} \ \
a_{\rm crit} \approx 130~{\rm au} \left( \frac{R_{\rm crit}}{{\rm km}} \right)^{-\frac{1}{2}}
,
\label{SnowCrit}
\end{equation}
\begin{equation}
{\rm (ROCK)} \  R_{\rm crit} \approx 890~{\rm km} 
\left(\frac{a_{\rm crit}}{{\rm au}}\right)^{-2}
\ \ {\rm or} \ \
a_{\rm crit} \approx 30~{\rm au} \left( \frac{R_{\rm crit}}{{\rm km}} \right)^{-\frac{1}{2}}
,
\label{RockCrit}
\end{equation}
\begin{equation}
{\rm (IRON)} \  R_{\rm crit} \approx 270~{\rm km} 
\left(\frac{a_{\rm crit}}{{\rm au}}\right)^{-2}
\ \ {\rm or} \ \
a_{\rm crit} \approx 16~{\rm au} \left( \frac{R_{\rm crit}}{{\rm km}} \right)^{-\frac{1}{2}}
.
\label{IronCrit}
\end{equation}

\noindent{}Eqs. (\ref{SnowCrit})-(\ref{IronCrit}) are conveniently compact relations that provide estimates which are broadly applicable across all $t_{\rm cool}$ for typical hot white dwarfs.

{\rev For other hot white dwarfs whose masses significantly differ from $0.6M_{\odot}$, we can assess the prospects of defining critical values of $R$ and $a$ by estimating functional forms from Fig.~\ref{Fig:AgeTemp}. All curves on that plot, aside from the $0.4M_{\odot}$ curves, are well estimated with $T_{\rm WD} \sim t_{\rm cool}^{p}$ where $p = \left\lbrace -0.30, -0.20 \right\rbrace$. Given that $R_{\rm i}a^2$ is completely independent of $t_{\rm cool}$ when $p=-0.25$, this range of $p$ is sufficiently small and appropriately centred in order for us to conclude that this useful coincidence occurs for the majority of currently observed single white dwarfs.}

\subsection{Linking fragmentation and sublimation}

Although we have now established that sublimation occurs quickly around hot white dwarfs, the timescale is not necessarily quick enough to occur within a single orbit (as modelled in, e.g. \citealt{broetal2017,mcdver2021}). We can quantify this notion by considering the following ratio

\begin{equation}
\frac{t_{\rm sub}}{P}
=
\frac{
2 R \rho \mathcal{L} \sqrt{GM_{\rm WD}a\left(1-e^2\right)}
}
{\pi \sigma R_{\rm WD}^2 T_{\rm WD}^4}
= 
\frac{
2 R \rho \mathcal{L} \sqrt{GM_{\rm WD}q\left(2-\frac{q}{a}\right)}
}
{\pi \sigma R_{\rm WD}^2 T_{\rm WD}^4},
\label{PerEq}
\end{equation}

\noindent{}where $P$ is the object's orbital period. We have written the equation in terms of the orbital pericentre $q$ because that is often the more reported and physically relatable quantity (rather than $e$) when discussing asteroids that are kicked on highly eccentric orbits around white dwarfs. For such highly eccentric orbits, Eq. (\ref{PerEq}) illustrates that $t_{\rm sub}/P$ is nearly independent of $a$, and depends on the pericentre only as $\approx \sqrt{q}$. 

We plot Eq. (\ref{PerEq}) in Fig. \ref{Fig:NumOrbits}. The top panel displays curves for a $1$~m boulder at two different $q$ values of $2R_{\odot}$ and $1$~au (the choice of $a$ has little effect on the results, but is assumed to be 100~au). This plot illustrates that around hot white dwarfs, only rocky and iron boulders (not snowy boulders) could survive for longer than one orbit, and only where $q > 2R_{\odot}$. The plot also illustrates that the survival rate for all objects is about 250 more orbits around a $T_{\rm WD}=20$,$000$~K white dwarf than a $T_{\rm WD}=80$,$000$~K white dwarf.

The bottom panel displays curves as a function of object radius $R$, such that all curves assume $q=2R_{\odot}$. This plot reinforces the top panel while providing perspective about where an object is expected to fragment. Eqs. (28-29) of \cite{mcdver2021} present an approximate prescription for tidal fragmentation that is simple enough for our purposes\footnote{This prescription neglects self-gravity, whose incorporation would increase the critical radius for tidal disruption.}:
~
\begin{equation}
R \gtrsim \sqrt{\frac{3R_{\rm WD}^3S}{8GM_{\rm WD}\rho}} 
\left( \frac{r}{R_{\rm WD}} \right)^{3/2}
=
\sqrt{\frac{3r^3S}{8GM_{\rm WD}\rho}}
\ge
\sqrt{\frac{3q^3S}{8GM_{\rm WD}\rho}}
,
\label{FragLim}
\end{equation}
\noindent{}where $S$ is the tensile strength of the object, assumed to be $S=10^3$~Pa for snow, $S=10^6$~Pa for rock, and $S=10^8$~Pa for iron. These critical fragmentation radii for snow and iron are indicated on the plot with vertical lines, corresponding to about 5~km and 400~km, respectively.

If an object with $R>5$~km fragments, then the maximum size of the resulting debris would be about 1~km \citep{steetal2026}. In that case, the sublimation time would be effectively reset to a lower value than for the parent body. Around hot white dwarfs the fragments would all completely sublimate in $\sim10^1-10^5$ orbits, or $0.013$~Myr-0.13~Gyr, depending on the material properties of the fragment. This range is wide enough to not exclude the possibility of an early dynamical instability manifesting itself observationally after the white dwarf cools.

\section{Discussion}

Our results help to quantify the dearth of observations of solid material around hot and young white dwarfs. 

{\rev These results reinforce the idea that sublimation of spherical objects occurs quickly around these stars. However, aspherical objects may be destroyed even more quickly through the SYORP, or the sublimative-YORP effect \citep{stejac2016,safetal2021}. \cite{verassteckloff2026} modelled SYORP destruction timescales for some 10~Myr white dwarfs; they showed that depending on the physical parameters of the object chosen, the SYORP destuction timescale is within about an order of magnitude either above or below those computed here in Figs. \ref{Fig:Cont1}-\ref{Fig:Cont2} for snow. However, for rock or iron, direct sublimation is usually much quicker than SYORP, especially beyond a few Roche radii from the white dwarf. The radiative YORP effect destroys objects on a timescale orders of magnitude longer than the SYORP effect.}

Our focus here has been on sublimation. However, gas may also be generated through collisions or atmospheric evaporation. We speculate that collisions do not play a significant role around hot white dwarfs. Given how relatively ineffectual radiative drag is (Section 3.1), and how strong sublimative effects are within $10R_{\odot}$ (Section 3.2), the scope for collisional cascades \citep{kenbro2017a,kenbro2017b,lietal2021,broetal2022} to sustain themselves before sublimation dominates is limited. As such, in this regime, the white dwarf Roche radius should not provide a meaningful demarcation between different types of pebbles, dust or gas except in the immediate aftermath of tidal disruption of a large asteroid or planet \citep{debetal2012,verleibonetal2014,vercarleietal2017,duvetal2020,malper2020a,malper2020b,kuretal2024}.

Atmospheric evaporation of planets orbiting white dwarfs as a source of gas generation \citep{schetal2019,galetal2024} is a relevant consideration given the extended gas disc around the (young and hot) 13.3~Myr-old white dwarf WD~J0914+1914 \citep{ganetal2019}. This gas has been inferred to arise from the evaporation of an ice giant atmosphere, with the ice giant residing at a location of roughly $0.07$~au$\approx$~$15R_{\odot}$. An ice giant has a complex internal structure and is not subject to complete sublimation, but at this location may be subject to fragmentation, at least according to Eq. (\ref{FragLim}). In fact, it has been speculated that this planet is not intact based on the mode of its speedy arrival at this location \citep{verful2019,verful2020}. Regardless, it has generated the observed gas.

This planet, along with the second-generation planet around HS~0209+0832 (Williams, G\"{a}nsicke, et al., In Press 2026), arguably accounts for about 30~per\,cent of all known {\rev (seven)} planets around white dwarfs \citep{thoetal1993,sigetal2003,luhetal2011,vanderburgetal2020,blaetal2021,limetal2024,zhaetal2024}. {\rev Although the numbers are small, this early high success rate provides motivation for further searches around very hot white dwarfs. Further, if more planets are found around these stars, we would not necessarily expect that the planets would orbit metal-enriched white dwarfs because} only a very small minority of metal-enriched white dwarfs have $T_{\rm WD} > 20,$$000$~K \citep[Fig. 3 of][]{wiletal2024}.

Obtaining evidence of solid bodies of any size orbiting hot white dwarfs can help constrain the models here. The critical parking locations $a_{\rm crit}$ of km-sized rocky and iron asteroids (Section 3.3) exceed the asymptotic giant branch engulfment radius \citep{musvil2012,madetal2016}, and so could easily explain pollution after the white dwarf cools through delayed gravitational instability\footnote{Stability boundaries may be crossed during giant branch mass loss, but the resulting collisions or ejections may not manifest until later, at different white dwarf cooling ages \citep{debetal2012,vermusbonetal2013,musetal2014,vergan2015,hampor2016,vermusganetal2016,petmun2017,steetal2017,musetal2018,maletal2020a,maletal2020b,maldonadoetal2021,maldonadoetal2022,vergeomusetal2021,lietal2022,ocoetal2022,trietal2022,verros2023}.}. However, larger rocky asteroids of, e.g., 100-1,000km in size, can be perturbed on highly eccentric orbits around newly-born white dwarfs and still survive for tens-to-hundreds of Myr, representing an important relic of a time when observations are difficult to make.

Also, the values of $a_{\rm crit}$ have other implications. For both iron and rock, $a_{\rm crit}$ exceeds the distance where the solar system's Main Belt of asteroids will reach around the solar white dwarf. The implication is then that exo-Kuiper Belt analogues are much more important for eventually polluting white dwarfs than closer-in material. Further, the much higher value of $a_{\rm crit}$ for snow highlights how observations of water-rich pollution signatures likely arose from progenitor asteroids that were large enough to retain volatiles in their interior but with surface layers that resisted sublimation \citep{malper2016,malper2017a,malper2017b}. Alternatively, these progenitors could have primarily arisen from exo-Oort clouds \citep{alcetal1986,vershagan2014,stoetal2015,caihey2017,ocoetal2023b,pharei2024,veras2025}.

\section{Summary}

White dwarf planetary systems endure a hot ($T_{\rm WD}\gtrsim 20$,$000$) and nascent ($t_{\rm cool}\approx0$-$20$~Myr) post-nebular phase that is largely hidden from observations, but crucially sets the course of subsequent evolution for the next 10~Gyr and beyond. In this paper, we have characterised the survival of solid bodies and distribution of sublimated material during this early epoch of white dwarf cooling, and have demonstrated the following:

\begin{itemize}

\item Despite high white dwarf temperatures, the radiative drag of dust, pebbles and boulders is a non-factor when compared to the speed of sublimation (Eqs. \ref{tPRmax}-\ref{tdest}). Consequently, both processes may be decoupled.

\item Most dust, pebbles or asteroids smaller than about 1~km will be sublimed within a few white dwarf Roche radii on timescales of years (Fig. \ref{Fig:Cont1}).

\item Metre-scale snowy objects are easily sublimed on au-scales within decades (Fig. \ref{Fig:Cont2}).

\item The critical size and distance of an object that can survive this hot post-nebular phase for long enough to later become a metal-enricher (or polluter) of white dwarfs is given by Eqs. (\ref{SnowCrit})-(\ref{IronCrit}). The simplicity and wide applicability of these (order of magnitude) equations arise because of a coincidental empirical dependence of $t_{\rm cool}$ on $T_{\rm WD}$ during the hottest phase of typical white dwarfs.

\item Objects which drift into this ``parked'' zone due to giant branch mass loss may be perturbed on a highly eccentric orbit with a pericentre close to (within a few $R_{\odot}$) a hot white dwarf at any cooling age. If this perturbation occurs when $t_{\rm cool}\lesssim20$~Myr, then the object is not necessarily doomed to a gaseous demise. There is a restricted region of parameter space where the object may survive intact on Gyr timescales, and another where the object is fragmented, but the largest km-scale fragments survive until $T_{\rm WD}<20,$$000$~K (Fig. \ref{Fig:NumOrbits}).

\item The critical parked distances from Eqs. (\ref{SnowCrit})-(\ref{IronCrit}) emphasise the greater importance of exo-Kuiper Belts and exo-Oort clouds rather than exo-Main Belt analogues when explaining metal-enrichment or pollution around white dwarfs cooler than 20,000~K.

\end{itemize}

\section*{Acknowledgements}

{\rev We thank the reviewer for helpful comments which have improved the manuscript.} All authors were supported by Royal Society grant IEC\textbackslash R3\textbackslash 243043 and JSPS grant JPJSBP120255708. We also thank Antoine B\'{e}dard for useful discussions.

\section*{Data Availability}

All data presented in this paper is available upon reasonable request to the authors.


\bibliographystyle{mnras}
\bibliography{DVbibfile13}

\bsp
\label{lastpage}
\end{document}